\documentclass[aps,pra,reprint,onecolumn,notitlepage,eqsecnum,showkeys]{revtex4-2}

\usepackage{hyperref}

\newcommand{\bq}{\begin{eqnarray}}
\newcommand{\eq}{\end{eqnarray}}
\newcommand{\rr}{{\bf r}}
\newcommand{\kk}{{\bf k}}
\newcommand{\xx}{{\bf x}}
\newcommand{\nn}{{\bf n}}
\newcommand{\rrp}{{\bf r}^\prime}

\begin{document}
\title{Static screening in a degenerate electron plasma}

\author{Riccardo Fantoni}
\email{riccardo.fantoni@scuola.istruzione.it}
\affiliation{Universit\`a di Trieste, Dipartimento di Fisica, strada
  Costiera 11, 34151 Grignano (Trieste), Italy}

\date{\today}

\begin{abstract}
We present a self contained derivation of the Friedel oscillations in a 
degenerate ideal electron plasma using a not commonly known theorem on the asymptotic 
behavior of the Fourier transform of a generalized function presenting some 
singularities. 
\end{abstract}


\maketitle

\section{Introduction}
An electron gas is a system of identical point-like charged fermions, of mass and 
charge those of the electron, neutralized by a uniform, inert background of opposite 
charge.

Some recent studies on the electron gas or {\sl the jellium} are about two dimensional 
systems \cite{Fantoni95a,Fantoni95b,Fantoni03a,Fantoni08c,Fantoni12b,Fantoni18c,Fantoni19a,Fantoni23a} or three dimensional ones 
\cite{Fantoni95b,Fantoni13g,Fantoni16b,Fantoni18a,Fantoni21b,Fantoni21d,Fantoni21i}. Here we will just consider a 
degenerate ideal electron gas in three dimensions.

We present a self contained derivation of the Lindhard theory of static screening in 
a degenerate ideal electron plasma which explains the nature of the Friedel 
oscillations. We follow Section 4.1 of the book ``Coulomb Liquids'' of N. H. March and 
M. P. Tosi \cite{March-Tosi}. But in the end we use a not commonly known theorem on the 
asymptotic behavior of the Fourier transform of a generalized function presenting some 
singularities.  

\section{A simple derivation}
Suppose we switch on an appropriately screened test charge potential
$\delta V$ (actually the so called Hartree potential) in a uniform ideal Fermi
gas. The Hartree potential $\delta V(r)$ created at a distance r
from a static point charge of magnitude $e$ should be evaluated 
self-consistently from the Poisson equation,
\bq
{\bf \nabla}^2\delta V(r)=-4\pi e^2[\delta(\rr)+\delta n(r)]~~, 
\eq
where $\delta(\rr)$ ia a Dirac delta function in three dimensions and $\delta n(r)$ 
is the change in electronic density induced by the foreign charge. As usual we will 
adopt the notation of indicating in bold the vectors so that $r=|\rr|$ is the modulus 
of the three dimensional position vector. The electron density 
$n(\rr)$ may be written as 
\bq \label{eq:n}
n(\rr)=2\sum_\kk |\psi_\kk(\rr)|^2~~,
\eq
where $\psi_\kk(\rr)$ are single-electron orbitals, the sum over $\kk$
is restricted to occupied orbitals ($|\kk|\leq k_F$, $k_F$ Fermi wave vector)
and the factor 2 comes from the sum over spin orientations and is needed for 
the paramagnetic state (equal population of up and down spins) taken under exam. 
We must now calculate how the orbitals in
the presence of the foreign charge, differ from plane waves 
$\exp(i\kk\cdot\rr)$. We use for this purpose the Schr\"odinger equation,
\bq \label{eq:Se}
{\bf \nabla}^2\psi_\kk(\rr)+\left[k^2-\frac{2m}{\hbar^2}\delta V(r)\right]
\psi_\kk(\rr)=0~~,
\eq  
having imposed that the orbitals reduce to plane waves with energy 
$\hbar^2 k^2/(2m)$ at large distance
\footnote{This approach (which lead to the Random Phase Approximation, RPA)
is approximate insofar as the potential entering the {Schr\"odinger} equation
has been taken as the Hartree potential, thus neglecting exchange and 
correlation between an incoming electron and the electronic screening cloud.}.

With the aforementioned boundary condition the Schr\"odinger equation
may be converted into an integral equation,
\bq \label{eq:psi}
\psi_\kk (\rr)=\frac{1}{\sqrt{\Omega}} e^{i\kk\cdot\rr}+\frac{2m}{\hbar^2}
\int G_k(|\rr-\rrp|)\delta V(r')\psi_\kk(\rrp) d\rrp~~,
\eq
where $\Omega$ is the volume of the system and we have the 
``spherical wave'' solution $G_k(r)=-\exp(ikr)/(4\pi r)$. 
Here we used the property $\nabla^2v(r)=-4\pi\delta(\rr)$ for the 
Coulomb potential $v(r)=1/r$ in three dimensions, or in Fourier $q$-space 
$q^2v(q)=4\pi$. So that in the $k\to 0$ limit $G_k(r)$ reduces to 
$-v(r)/(4\pi)$. And we used the property of the Fourier transform to 
change a convolution into a product. After all, note that
$\nabla^2G_k(r)=-k^2G_k(r)+\delta(\rr)$, or in Fourier $q$-space
$(k^2-q^2)G_k(q)=1$.

Within linear response theory we can replace $\psi_\kk(\rr)$ by
$\exp(i\kk\cdot\rr)/\sqrt{\Omega}$ inside the integral. This yields 
(see Appendix \ref{app2})
\bq
\delta n(r)=-\frac{mk_F^2}{2\pi^3\hbar^2}\int j_1(2k_F|\rr-\rrp|) 
\frac{\delta V(r')}{|\rr-\rrp|^2}d\rrp~~,
\eq 
with $j_1(x)$ being the first-order spherical Bessel function 
$[\sin(x)-x\cos(x)]/x^2$. Using this result in the Poisson equation
we get
\bq \label{eq:pe}
{\bf \nabla}^2\delta V(r)=-4\pi e^2\delta(\rr)+\frac{2mk_F^2 e^2}
{\pi^2\hbar^2}\int j_1(2k_F|\rr-\rrp|)\frac{\delta V(r')}{|\rr-\rrp|^2}
d\rrp~~,
\eq 
which is easily soluble in Fourier transform (see Appendix \ref{app1}). 
Writing $\delta V(k)=4\pi e^2/[k^2\varepsilon(k)]$ we find,
\bq \label{eq:df}
\varepsilon(k)=1+\frac{2mk_Fe^2}{\pi\hbar^2 k^2}\left[1+\frac{k_F}{k}
\left(\frac{k^2}{4k_F^2}-1\right)\ln\left|\frac{k-2k_F}{k+2k_F}\right|
\right]~~,
\eq
which is the static dielectric function in RPA.

For $k\rightarrow 0$ this expression gives $\varepsilon(k)\rightarrow
1+k_{TF}^2/k^2$ with $k_{TF}=3\omega_p^2/v_F^2$ ($\omega_p$ being the 
plasma frequency and $v_F$ the Fermi velocity.) i.e. the result of the
Thomas-Fermi theory. However $\varepsilon(k)$ has a singularity at 
$k=\pm 2k_F$, where its derivative diverges logarithmically
\footnote{The discontinuity in the momentum distribution across the Fermi
surface introduces a singularity in elastic scattering processes with 
momentum transfer equal to $2k_F$.}.
This singularity in $\delta V(k)$ determines, after Fourier transform, the 
behaviour of $\delta V(r)$ at large $r$. $\delta V(r)$ turns out to be 
an oscillating function \cite{Friedel1958}
rather than a monotonically decreasing function as in the Thomas-Fermi 
theory. Indeed,
\bq\nonumber
\delta V(r)&=&\int\frac{d\kk}{(2\pi)^3}\frac{4\pi e^2}{k^2\varepsilon(k)}
e^{i\kk\cdot\rr}\\\nonumber
&=&\int_0^\infty k^2dk\int_0^\pi\sin\theta\,d\theta\int_0^{2\pi}
d\varphi\,\frac{4\pi e^2}{(2\pi)^3k^2\varepsilon(k)}e^{ikr\cos\theta}\\
\nonumber
&=&\frac{e^2}{\pi}\int_0^\infty dk\int_{-1}^1 d(\cos\theta)\,
\frac{e^{ikr\cos\theta}}{\varepsilon(k)}\\\nonumber
&=&\frac{e^2}{i\pi r}\int_0^\infty dk\,
\frac{e^{ikr}-e^{-ikr}}{k\varepsilon(k)}\\
&=&\frac{e^2}{i\pi r}\int_{-\infty}^\infty dk\frac{e^{ikr}}
{k\varepsilon(k)}~~,
\eq
where we expressed the three dimensional integral in 
$d\kk=(dk)(k\,d\theta)(k\sin\theta\,d\varphi)$ with $k=|\kk|\in [0,\infty]$,
$\theta\in [0,\pi]$, and $\varphi\in [0,2\pi]$ and we used the fact that  
$\varepsilon(k)$ is an even function. The integrand has non-analytic behaviour at 
$k=\pm 2k_F$,
\bq
\left[\frac{1}{k\varepsilon(k)}\right]_{k\rightarrow\pm 2k_F}=
-A(k-(\pm)2k_F)\ln|k-(\pm)2k_F|+\mbox{regular terms}~~,
\eq
with $A=B/(B+4k_F^2)^2$ where $B=2mk_Fe^2/(\pi\hbar^2)=k_{TF}^2/2$. Hence,
\bq
\nonumber
&&\delta V(r)|_{r\rightarrow \infty}\\\nonumber
&&=-\frac{Ae^2}{i\pi r}\int_{-\infty}
^\infty dk\,e^{ikr}[(k-2k_F) \ln|k-2k_F|-(k+2k_F) \ln|k+2k_F|]\\\nonumber
&&=-\frac{2Ae^2}{\pi r}\lim_{a\to0^+}\int_0^\infty dk\,e^{-ak}\sin(kr)[(k-2k_F) \ln|k-2k_F|-(k+2k_F) \ln|k+2k_F|]\\\nonumber
&&=2Ae^2\left\{\ln(2k_F)\frac{4k_F}{\pi r^2}+\frac{\cos(2k_Fr)}{r^3}+2\frac{\cos(2k_Fr){\cal I}m[E_1(i2k_F r)]}{\pi r^3}+2\frac{\sin(2k_Fr){\cal R}e[E_1(i2k_F r)]}{\pi r^3}\right\}\\\label{eq:fo}
&&=2Ae^2\left\{\ln(2k_F)\frac{4k_F}{\pi r^2}+\frac{\cos(2k_Fr)}{r^3}-\frac{1}{\pi k_Fr^4}+{\rm O}(1/r^5)\right\},
\eq
where $E_n(z)$ is the exponential integral function. 
This result is based on a theorem on Fourier transforms 
(see theorem 19 in \cite{Lighthill}), stating that the asymptotic
behaviour of $\delta V(r)$ is determined by the low-$k$ behaviour as well as 
by the singularities of $\delta V(k)$, i.e. the points where it is not analytic. 
Obviously, in the present case, the 
asymptotic contribution from the singularities is dominant over the 
exponential decay of Thomas-Fermi type, due to the analytic part of the 
Fourier transform. The result (\ref{eq:fo}) implies that the screened
ion-ion interaction in a metal has oscillatory character and ranges over 
several shells of neighbours.

\section{Conclusions}
We presented a self contained derivation of the Lindhard theory of static screening 
in a degenerate ideal electron plasma which explains the nature of the Friedel 
oscillations. This derivation can be used in statistical physics books for graduate 
students. We followed Section 4.1 of the book ``Coulomb Liquids'' of N. H. March 
and M. P. Tosi \cite{March-Tosi}.

\appendix
\section{From Eqs. (\ref{eq:n}), (\ref{eq:psi}) to Eq. (\ref{eq:pe})}
\label{app2}
Using periodic boundary conditions on the box of volume $\Omega=L^3$ containing the 
plasma we conclude that $\kk=(2\pi/L)\nn$ where $\nn$ is a triplet of integers.
Therefore $(1/\Omega)\sum_\kk\ldots\to\int_{|\kk|<k_F}d\kk/(2\pi)^3\ldots$. Now using 
Eq. (\ref{eq:psi}) into Eq. (\ref{eq:n}) we find
\bq
n(\rr)&=&\frac{2}{\Omega}\sum_\kk\left\{1+\frac{2m}{\hbar^2}\int\delta V(r')
2{\cal R}e\left[G_k(|\rr-\rr'|)e^{i\kk\cdot(\rr'-\rr)}\right]\,d\rr'+\ldots\right\},
\eq
where we omitted terms of order $(\delta V)^2$. Therefore we find
\bq\nonumber
\delta n(r)&=&\frac{4m}{\hbar^2\Omega}\sum_\kk\int\delta V(r')2{\cal R}e\left[G_k(|\rr-\rr'|)e^{i\kk\cdot(\rr'-\rr)}\right]\,d\rr'\\\nonumber
&=&-\frac{4m}{\hbar^2}\int d\rr'\,\frac{\delta V(r')}{4\pi|\rr-\rr'|}2{\cal R}e\int_{|\kk|<k_F}\frac{d\kk}{(2\pi)^3}e^{ik|\rr-\rr'|}e^{i\kk\cdot(\rr'-\rr)}\\
&=&-\frac{mk_F^2}{2\pi^3\hbar^2}\int j_1(2k_F|\rr-\rr'|)\frac{\delta V(r')}{|\rr-\rr'|^2}\,d\rr',
\eq
where we used 
\bq
\int_{|\kk|<k_F} d\kk\, e^{ikr}e^{i\kk\cdot\rr}=
2\pi i\frac{1+2k_F^2r^2+e^{i2k_Fr}(-1+i2k_Fr)}{4r^3}.
\eq

\section{Derivation of the static dielectric function}
\label{app1}
The Fourier transform of Eq. (\ref{eq:pe}) gives
\bq
-k^2\delta V(k)=-4\pi e^2+\frac{k_Fme^2}{\pi^2\hbar^2}I(\tilde{k})\delta V(k),
\eq
where $\delta V(k)=\int e^{i\kk\cdot\rr}\delta V(r)\,d\rr$ and we used the property of 
the Fourier transform to change a convolution into a product to find
\bq
I(\tilde{k})=\int\frac{j_1(x)}{x^2}e^{i\tilde{\kk}\cdot\xx}\,d\xx,
\eq
where $\tilde{\kk}=\kk/2k_F$ and the integration is over the whole three dimensional 
space so that $d\xx=x^2\,dx\sin\theta\,d\theta\,d\varphi$. Since 
$\tilde{\kk}\cdot\xx=\tilde{k}x\cos\theta$ and $x=|\xx|\in [0,\infty]$, 
$\theta\in [0,\pi]$, $\varphi\in [0,2\pi]$, we find
\bq
I(\tilde{k})=4\pi\int_0^\infty\frac{j_1(x)}{\tilde{k}x}\sin(\tilde{k}x)\,dx
=2\pi\left[1-\frac{\tilde{k}^2-1}{\tilde{k}}{\rm arctanh}(\tilde{k})\right].
\eq
Recognizing that
\bq
{\rm arctanh}(\tilde{k})=\frac{1}{2}\ln
\left|\frac{\tilde{k}+1}{\tilde{k}-1}\right|,
\eq
one readily finds Eq. (\ref{eq:df}).

\bibliography{tosi}

\end{document}